\documentclass[12pt,a4paper,reprint]{revtex4-1}
\usepackage{amsmath,amssymb}
\usepackage[utf8]{inputenc}
\usepackage{graphicx}
\usepackage{color}
\usepackage[english]{babel}
\usepackage[encapsulated]{CJK}
\begin{document}

\begin{CJK*}{UTF8}{}

\title{Transport of Sputtered Particles in Capacitive Sputter Sources}
\author{Jan Trieschmann}
\author{Thomas Mussenbrock}
\affiliation{Ruhr University Bochum, Department of 
Electrical Engineering and Information Sciences, 
Institute of Theoretical Electrical Engineering, 
D-44780 Bochum, Germany}
\date{\today}
\thanks{Copyright 2015 American Institute of Physics. This article may be downloaded for personal use only. Any other use requires prior permission of the author and the American Institute of Physics. The following article appeared in Journal of Applied Physics \textbf{118}, 033302 (2015) and may be found at http://scitation.aip.org/content/aip/journal/jap/118/3/10.1063/\\1.4926878.}

\begin{abstract}
\noindent The transport of sputtered aluminum inside a multi frequency capacitively coupled plasma chamber is simulated by means of a kinetic test multi-particle approach. A novel consistent set of scattering parameters obtained for a modified variable hard sphere collision model is presented for both argon and aluminum. An angular dependent Thompson energy distribution is fitted to results from Monte Carlo simulations and used for the kinetic simulation of the transport of sputtered aluminum. For the proposed configuration the transport of sputtered particles is characterized under typical process conditions at a gas pressure of $p=0.5$~Pa. It is found that -- due to the peculiar geometric conditions -- the transport can be understood in a one dimensional picture, governed by the interaction of the imposed and backscattered particle fluxes. It is shown that the precise geometric features play an important role only in proximity to the electrode edges, where the effect of backscattering from the outside chamber volume becomes the governing mechanism.
\end{abstract}

\maketitle

\end{CJK*}

\newpage

\section{Introduction}\label{sec:introduction}

\noindent Capacitively coupled plasmas are commonly used for plasma enhanced chemical vapor deposition (PECVD) applications \cite{lieberman_principles_2005}. As recently demonstrated by Bienholz et al.\ \cite{bienholz_dual_2012, bienholz_multiple_2013, bienholz_electrical_2014, bienholz_kapazitiv_2014} capacitive discharges operated at multiple driving frequencies can be also employed for physical vapor deposition (PVD) and sputtering applications for the deposition of metallic, ceramic, and/or magnetic thin films. In contrast to cathodic arc evaporation \cite{anders_review_2014}, magnetically enhanced PVD processes such as the widely used direct current magnetron sputtering (dcMS) \cite{maniv_aspects_1983}, or the recently technologically evolving high power impulse magnetron sputtering (HiPIMS or HPPMS) \cite{anders_review_2014, gudmundsson_high_2012}, multi-frequency capacitively coupled plasmas (MFCCP) for sputtering applications have the advantage of a diverse choice of materials, including non-conducting and ferromagnetic target materials \cite{bienholz_dual_2012}. For these materials DC driving sources and/or magnetic field enhancement may not be a suitable option. Despite the fact that the MFCCP reactor proposed by Bienholz et al.\ has been experimentally characterized in great detail, so far only a preliminary explanation of the transport phenomena of sputtered particles is provided \cite{bienholz_multiple_2013}. For the interpretation of some of the experimental findings there is a lack of reliable simulation data.

To close the gap between experiments and theory the transport of sputtered particles is studied under realistic discharge conditions. In order to characterize the geometrical constraints for the deposition behavior the transport of sputtered aluminum is first simulated within a fully three dimensional geometry. The interaction of the sputtered aluminum with both the background argon gas and the backscattered aluminum is subsequently analyzed using a reduced one dimensional setup. The comparison of the two models reveals the suitability and limitations of the reduced 1D model governing the transport and thus the density and mean velocity of the sputtered particles.

For the proposed theoretical approach a strongly modified version of the direct simulation Monte Carlo (DSMC) code dsmcFoam \cite{scanlon_open_2010} is used. (dsmcFoam is part of the open source software package OpenFOAM \cite{_openfoam:_2014}). Although the code in principle allows to self-consistently simulate the background flow field as well as the sputtered particle transport, in this work, for the sake of computational feasibility, only the transport of sputtered particles is simulated using a variant of the test particle method (TPM). This is necessary because the sputtered aluminum particles constitute a trace minority in the overall gas composition, mainly the argon background. The TPM has been used for numerous studies of sputtering processes as reported by a number of researchers and research groups \cite{somekh_thermalization_1984, motohiro_monte_1984, turner_monte_1989, myers_monte_1992, kersch_selfconsistent_1994, serikov_monte_1996, clenet_experimental_1999, kadlec_simulation_2007, van_aeken_metal_2008, depla_magnetron_2012, jimenez_comprehensive_2012, nanbu_synthetic_2013, lundin_tiar_2013}. Several different implementations have been referenced in the literature, for example a single particle approach \cite{somekh_thermalization_1984}, or a treatment of independent particles in a parallel fashion following the test multi-particle method (TMPM) \cite{kersch_selfconsistent_1994, serikov_monte_1996}. In either case, three aspects are of particular importance: i) the details of the numerical implementation of the TPM itself, ii) the underlying collision model for the particle interaction, and iii) the angular and energy distribution of initially injected sputtered particles. All of these aspects are innovatively reconsidered in this work (cf., section~\ref{sec:numerical_method}).

The manuscript is organized as follows: After a brief description of the (experimental) discharge configuration and the operating conditions in section~\ref{sec:setup}, a review of the employed numerical scheme is presented in section~\ref{sec:numerical_method}. Particular emphasis is placed on the collisional interaction of sputtered particles with the background gas as well as on their initial energy distribution after sputtering occurs. Then in section~\ref{sec:results}, firstly, simulation results for the realistic three dimensional MFCCP vacuum chamber are discussed with the focus on the geometric constraints affecting the details of the particle transport. Secondly, simulation results from a reduced one dimensional model are discussed highlighting the fundamental transport mechanism. A parameter study of the density and the mean velocity profiles for the sputtered aluminum is presented. The nature of the governing transport process is discussed on the basis of the particles' velocity distribution functions (VDF). Finally, in section~\ref{sec:conclusion} the results are summarized and conclusions are drawn.

\section{Discharge Setup}\label{sec:setup}

\begin{figure}[t!]
	\centering
	\resizebox{8cm}{!}{\includegraphics{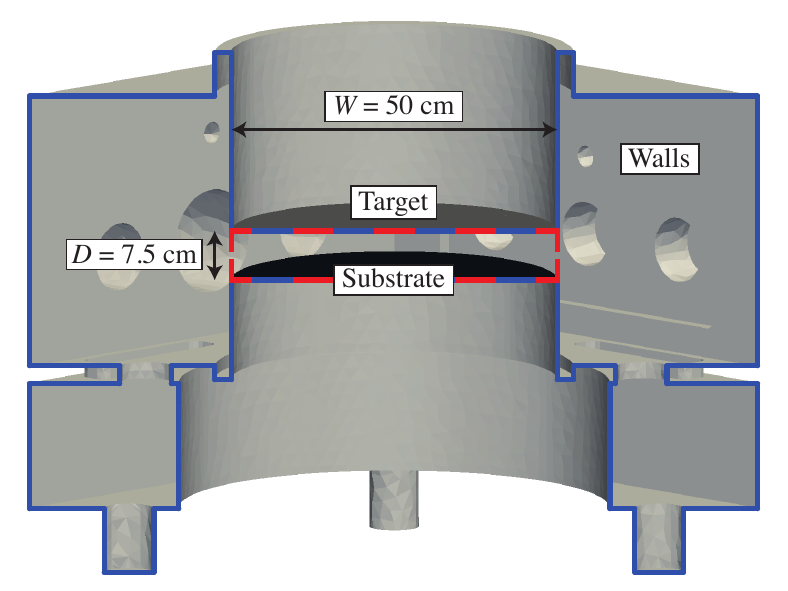}}
	\caption{Sectional drawing of the MFCCP vacuum 
	chamber. The driven top electrode is the target, 
	while a substrate can be mounted at the opposite 
	bottom electrode. The latter as well as the chamber
	walls are grounded. The red dashed frame indicates 
	the electrode gap region.}
	\label{fig:geometry}
\end{figure}

\noindent To obtain a principle understanding of the subject under investigation in this work, a brief review of the discharge setup previously described and investigated by Bienholz et al.\ is instructive \cite{bienholz_multiple_2013, bienholz_kapazitiv_2014}. A sectional drawing of the vacuum chamber is depicted in figure~\ref{fig:geometry}. The stainless steel vessel forms a cuboid of 80~cm side lengths and a height of 65~cm. Two opposing circular electrodes of diameter $W = 50$~cm are separated by a gap of distance $D = 7.5$~cm. While the upper (target) electrode is driven with multiple radio-frequencies (RF), the lower (substrate) electrode, as well as the remaining parts of the reactor chamber (i.e., the walls) are electrically grounded. Consequently, under operation this geometrically asymmetric discharge exhibits a substantial self-bias voltage. The discharge is typically simultaneously driven by three independent power supplies at frequencies of 13.56, 27.12 and 60 MHz. Taking advantage of the electrical asymmetry effect (EAE) \cite{heil_possibility_2008, czarnetzki_electrical_2009} the MFCCP has been demonstrated to be an ideal candidate as a capacitively coupled sputtering system. The target electrode is made of aluminum and is subject to undergo sputtering by ion bombardment. The remaining components are made of stainless steel. Since there is no necessity to magnetically enhance the plasma, besides the inherent plasma profile over the electrode no additional lateral inhomogeneity is introduced due to magnetic fields (e.g., a race-track). Consequently, advantage can be taken from a nearly homogeneous target utilization over a quite large area of $A_\textrm{target} \approx 1963~\textrm{cm}^2$.

Bienholz et al.\ provide a sophisticated experimental characterization based on a variety of diagnostic techniques (i.e., optical emission spectroscopy, Langmuir probe, and micro balance deposition measurements). They further specify the parameters for a reference scenario to be investigated in this work: In steady state operation the gas temperature $T = 650$~K is measured with a confidence interval of $\pm 50$~K. At a gas pressure of $p = 0.5$~Pa the background gas density (argon) then is $n_\textrm{Ar} = 5.572 \times 10^{13} \textrm{ cm}^{-3}$. The energy of monoenergetic argon ions impinging the target, $E_\textrm{i}$, is taken from the measured self-bias voltage $V_{\textrm{SB}} = 437$~V. The assumption of monoenergetic ions is justified because the ion plasma frequency $\omega_\textrm{pi} \approx 28 \times 10^6 \textrm{ s}^{-1}$ is significantly smaller than the lowest driving frequency $\omega_\textrm{rf, min} = 2 \pi \times 13.56 \textrm{ MHz} \approx 85.2 \times 10^6 \textrm{ s}^{-1}$. Ions therefore experience only the average electric field within the boundary sheath in front of the electrodes. Consequently they are not modulated. Moreover, under the assumption of a low collisionality (which is the case here), the ion energy distribution is not substantially broadened.

Advantage can be taken from simulations of the sputtering process itself using the code TRIDYN \cite{moller_tridyn_1984, hofsass_simulation_2014}. The angular and energy distributions of sputtered particles regarded in this work have been obtained by Bienholz et al.\ like so \cite{bienholz_multiple_2013}. The flux of aluminum particles sputtered from the target is estimated from the experiment as $\Gamma_\textrm{0} = 1.6 \times 10^{15} \textrm{cm}^{-2}\textrm{s}^{-1}$ \cite{bienholz_kapazitiv_2014}. A homogeneous flux of sputtered particles over the target surface can be assumed, because of the quite homogeneous plasma density in front. Moreover, based on retarding field measurements performed by Ries et al.\ \cite{ries_private_2015} resputtering from the substrate surface can be neglected. From the low ion energies of $E_\textrm{inc} \lesssim 50$~eV at the substrate the sputtering yield is estimated to be less than a few percent. Finally, at all surfaces complete adsorption of the impinging aluminum can be assumed (with a sticking coefficient $s=1$). Although rather simple, this is a good approximation for (low) energetic aluminum atoms ($\leq 25$~eV), regardless of their angular distribution \cite{hansen_atomistic_1999}.

Within the volume typical aluminum densities of approximately $10^{10} \textrm{~cm}^{-3}$ result. In comparison, the background density is on the order of $5 \times 10^{13} \textrm{~cm}^{-3}$. From the ratio of the two (i.e., $2 \times 10^{-4}$) it can be reasoned that aluminum is merely contained in traces. In consequence, even though the average energy of sputtered particles is roughly two orders of magnitude higher than the background's thermal energy ($k_\textrm{B} T \approx 56$~meV), gas heating and thus rarefaction of the background gas remain negligible. Effects like a sputtering wind \cite{hoffman_sputtering_1985} can therefore be excluded. (Note that the contrary is typically the case for HiPIMS. In that case a substantial rarefaction of the background may occur due to high metal densities comparable to the density of the process gas, paired with a high degree of ionization. The proposed model in the present form is inappropriate for this situation.)

In addition to the experimental analysis, Bienholz et al.\ also provide simulation results obtained from a one dimensional TPM code based on collisional data of Kuwata et al.\ \cite{kuwata_comparative_2003}. Despite the strong effort, these theoretical results do not seem to explain what is observed experimentally. In particular there are some quite contradictory discrepancies in the sputtered particle densities and fluxes. The goal of this work is to provide a detailed theoretical analysis of the underlying transport mechanism in order to fully understand and explain these seemingly contradictory results.

\section{Numerical Method}\label{sec:numerical_method}

\noindent As previously introduced, in this work a strongly modified version of the open-source DSMC solver dsmcFoam is employed \cite{scanlon_open_2010, _openfoam:_2014}. The original version of the package provides the basic functionality in terms of a Lagrangian particle motion (with one super-particle representing a given number of physical particles), binary collisions based on the variable hard sphere (VHS) model, as well as free-stream boundary conditions. Its capability for the simulation of the particle transport in sputtering systems is however limited. A number of necessary modifications have been applied to the code which comprise of the implementation of the M1 binary collision model \cite{morokoff_comparison_1998}, inlet and wall boundary conditions appropriate to accurately mimic sputtering from a given target surface assuming a modified Thompson distribution \cite{stepanova_estimates_2001}, and species selective particle sticking to the walls. The specifics of these modifications are detailed in subsections~\ref{sec:m1} and \ref{sec:thompson}.

For the transport simulation of trace species a simplified computational procedure can be adopted. The minority particles solely interact with a constant, non stationary background, but not among themselves. This (or an equivalent) assumption is common to test-particle methods in general. The background is then completely specified by its number density $n_\textrm{bg}$ and temperature $T$. The proposed TMPM model differs significantly from the previously described TMPM models \cite{kersch_selfconsistent_1994, serikov_monte_1996}. In our model a simplified no-time counter (NTC) method paired with a constant time-stepping scheme is used. For all super-particles (also referred to as simulators) equal weights are assumed. Correspondingly, the number of physical particles represented by one super-particle is $w=n V_\textrm{c} / N_\textrm{c}$, with the trace species number density $n$, the cell volume $V_\textrm{c}$ and the number of super-particles per cell $N_\textrm{c}$. The probability for a trace species simulator $i$ to undergo collision with the background can be specified as \cite{serikov_monte_1996}
\begin{align}
	P_\textrm{i,bg} = n_\textrm{bg} 
	\left( \sigma_\textrm{T} V_\textrm{r} 
	\right)_\textrm{i,bg} \Delta t,
	\label{eq:collision_probability_basic}
\end{align}
where $V_\textrm{r} = \left| \vec{v}_\textrm{i} - \vec{v}_\textrm{bg} \right|$ is the relative velocity between the simulator $i$ and the background sampled from a Maxwellian distribution with temperature $T$. $\sigma_\textrm{T} = \sigma_\textrm{T}(V_\textrm{r})$ is the velocity dependent total collision cross-section integrated up to a finite cut-off $b_\textrm{max}$ of the impact parameter $b$ (cf., section~\ref{sec:m1}). $\Delta t$ is the discrete simulation time step. In the fashion of the NTC method, the collision probability can be rewritten as \cite{bird_molecular_1994, kersch_transport_2011}
\begin{align}
	P_\textrm{i,bg} &= N_\textrm{cand} \left[ \dfrac{1}
	{N_\textrm{c}} \right] \left[ \dfrac{\left( 
	\sigma_\textrm{T} V_\textrm{r} \right)_\textrm{i,bg}}
	{\max\left( \sigma_\textrm{T} V_\textrm{r} 
	\right)_\textrm{i,bg}} \right], 
	\label{eq:collision_probability}
\end{align}
with the number of collision candidates per time step given by
\begin{align}
	N_\textrm{cand} = \left[ n_\textrm{bg} N_\textrm{c} 
	\max \left( \sigma_\textrm{T} V_\textrm{r} 
	\right)_\textrm{i,bg} \Delta t \right].
	\label{eq:collision_candidates}
\end{align}
The first pair of parentheses in eq.~(\ref{eq:collision_probability}) represents the probability of randomly drawing one out of $N_\textrm{c}$ simulators, the second pair of parentheses is a normalized collision probability for this choice. The computational efficiency can thus be largely improved by selecting only $N_\textrm{cand}$ candidates out of the $N_\textrm{c}$ simulators per cell. This representation is physically equivalent to the TMPM as described in \cite{kersch_selfconsistent_1994, serikov_monte_1996}, only that it is significantly more efficient. The restriction of a maximum allowed time step $\Delta t$ to ensure the probability of a collision to be $P_\textrm{col} \leq 1$, is not truly circumvented as also $N_\textrm{cand} \leq N_\textrm{c}$ has to be ensured. Thus $\Delta t \ll \tau_\textrm{c} $ is a priori chosen to be much smaller than the mean collision time $\tau_\textrm{c}$ \cite{kersch_transport_2011, sun_proper_2011}.

On collision, the actual scattering is evaluated \cite{bird_molecular_1994, kersch_transport_2011}: The pre-collision velocities are first transformed from the laboratory into the co-moving center of mass frame. In the co-moving frame local reference coordinates are then defined. Next, the impact parameter $b = b_\textrm{max}\sqrt{R_f}$ is selected (uniformly distributed over the cross-section) with a uniformly distributed random number $R_f \in \left[0,1\right]$ \cite{vincenti_introduction_1967, clenet_experimental_1999, van_aeken_metal_2008}. Along with the impact parameter also the scattering angle $\chi(b)$ is specified using in our case the M1 model (to be discussed next). The azimuthal angle $\psi = 2 \pi R_f$ is drawn (with another independent random number $R_f$) to completely specify the post-collision scattering direction in the local reference frame. The post-collision velocities are finally obtained in the laboratory frame (taking into account the rotation and change in coordinates).

\subsection{M1 Collision Parameters}\label{sec:m1}

\noindent In favor of the ordinary VHS model which comes with dsmcFoam the M1 collision model proposed by Morokoff and Kersch is used \cite{morokoff_comparison_1998}. (Therein an elaborated discussion of the M1 model is provided.) As reasoned, compared to the VHS model it gives a more accurate description of the collision dynamics at the cost of a slightly more sophisticated evaluation. This comes from the scattering direction being not isotropic on a sphere anymore. Despite that the M1 collision model can be used with essentially the same parameters as the VHS model (scaled by a factor of $\sqrt{4/3}$ \cite{kersch_selfconsistent_1994}), in this work a novel set of collisional parameters for all molecular interactions included in the simulation is proposed. This is necessary because for the collision processes of interest so far only rather inconsistent VHS or M1 data sets are available (in particular for metal species). In order to obtain a consistent set of cross-sectional data, the scattering (or deflection) angle \cite{vincenti_introduction_1967}
\begin{align}
	\chi(b,V_\textrm{r}) = \pi - 2 b 
	\int_{r_\textrm{A}}^{\infty} dr
	\left[ r^2 \sqrt{1 - \left(\dfrac{b}{r}\right)^2 
	-\dfrac{\Phi(r)}{\varepsilon(V_\textrm{r})}} 
	\right]^{-1} 
	\label{eq:exact_scattering_law}
\end{align}
is first calculated from an analytic interaction potential $\Phi(r)$ as a function of the impact parameter $b$ and the relative velocity $V_\textrm{r}$ (writing the collisional energy as $\varepsilon(V_\textrm{r}) = \frac{1}{2} m_\textrm{r} V_\textrm{r}^2$). The distance of closest approach is $r_\textrm{A}$ and the reduced mass is calculated from the individual particle masses $m_\textrm{r} = m_1 m_2 / (m_1 + m_2)$. The interaction potential is $\Phi(r) = A\exp{\left( - B r \right)}$ based on a Born-Mayer form \cite{born_zur_1932}. The well-documented set of parameters of Abrahamson is used \cite{abrahamson_born-mayer-type_1969}. The scattering dependence is obtained for energies $\varepsilon(V_\textrm{r})=\left\{0.01,~0.1,~1,~10,~100\right\}$~eV. Over the range of these energies the total deviation is minimized using the method of least squares. Correspondingly, the M1 model's linear relationship 
\begin{align}
	\chi(b,V_\textrm{r}) = \pi \left[ 1 - \left. 
	b \middle / d_\textrm{i}(V_\textrm{r}) \right. \right]
\end{align}
is used. The cut-off impact parameter $b_\textrm{max}$ equals $d_\textrm{i}(V_\textrm{r})$, the velocity dependent diameter of species $i$. Following the commonly used temperature power law for the viscosity $\mu(T) \propto \left( T / T_\textrm{ref} \right)^{\omega_\textrm{i}}$ \cite{vincenti_introduction_1967, bird_molecular_1994, kersch_transport_2011} the latter reads
\begin{align}
d_\textrm{i}(V_\textrm{r}) = d_\textrm{ref,i} 
\sqrt{\left( \dfrac{k_\textrm{B} 
T_\textrm{ref}}{\varepsilon(V_\textrm{r})} 
\right)^{\omega_\textrm{i}-1/2} \dfrac{1}{
\Gamma (5/2-\omega_\textrm{i})} }.
\end{align}

\begin{figure}[t]
	\centering
	\resizebox{8cm}{!}{
		\includegraphics{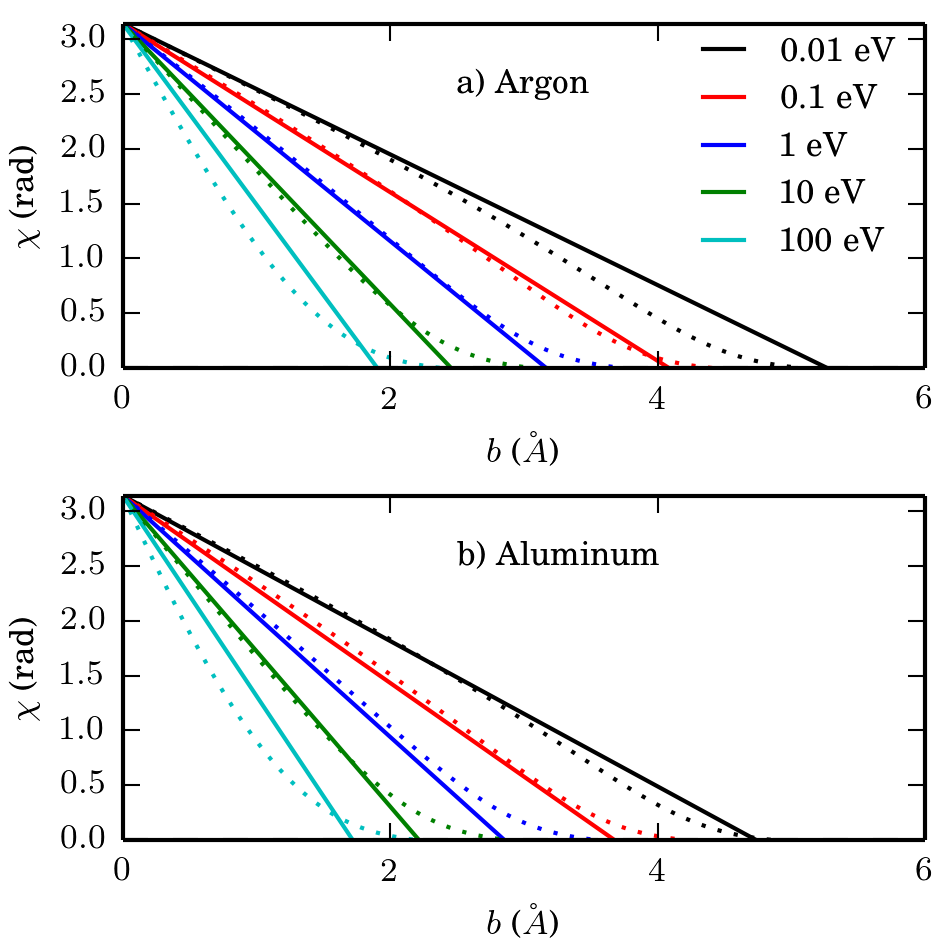}
	}
	\caption{Scattering angle $\chi$ plotted as a function of 
	the impact parameter $b$ for the linear dependence of 
	the M1 model (solid lines) and the analytic 
	calculations (dotted lines) for different interaction 
	energies. a) Argon (above) and b) aluminum (below).}
	\label{fig:impact_parameter_scattering}
\end{figure}

\begin{table}[t]
	\centering
	\caption{M1 collision parameters 
	obtained for argon and aluminum.}
	\label{tab:m1_parameters}
	\begin{tabular}{p{0.165\columnwidth} p{0.165\columnwidth} p{0.3\columnwidth} p{0.3\columnwidth}}
		\toprule
		& & Argon & Aluminum \\
		\hline
		$d_\textrm{ref,i}$ & $(\textrm{\AA})$ & 4.614 & 4.151 \\
		$\omega_\textrm{i}$ & $(-)$ & 0.721 & 0.72 \\
		\hline
	\end{tabular}
\end{table}

Figure~\ref{fig:impact_parameter_scattering} shows the obtained scattering angle $\chi(b,\varepsilon(V_\textrm{r}))$, as well as the linear approximation for argon (Fig.~\ref{fig:impact_parameter_scattering}a) and aluminum (Fig.~\ref{fig:impact_parameter_scattering}b). The obtained M1 parameters $d_\textrm{ref,i}$ and $\omega_\textrm{i}$ are listed in table~\ref{tab:m1_parameters}. Clearly, the linear fits resemble the interaction kinetics of the much more complex interaction potential reasonably well. For all energies the overall functional shape is similar. A systematic deviation is observed for impact parameters $b > d_\textrm{i}(V_\textrm{r})$ due to the assumption of a finite distance of molecular interaction. For $b \leq d_\textrm{i}(V_\textrm{r})$ the dependence of the scattering angle on the interaction energy is captured appropriately over a wide range of energies. Only for very low ($\varepsilon \leq 10$~meV) and quite high ($\varepsilon \geq 100$~eV) energies the M1 model more substantially deviates from the analytic calculations. This is a limitation of the power law assumption. Depending on the energy range of interest a more appropriate energy dependence may be chosen, possibly involving a more focused consideration of the kinetics of molecular interaction.

\begin{figure}[t]
	\centering
	\resizebox{8cm}{!}{
		\includegraphics{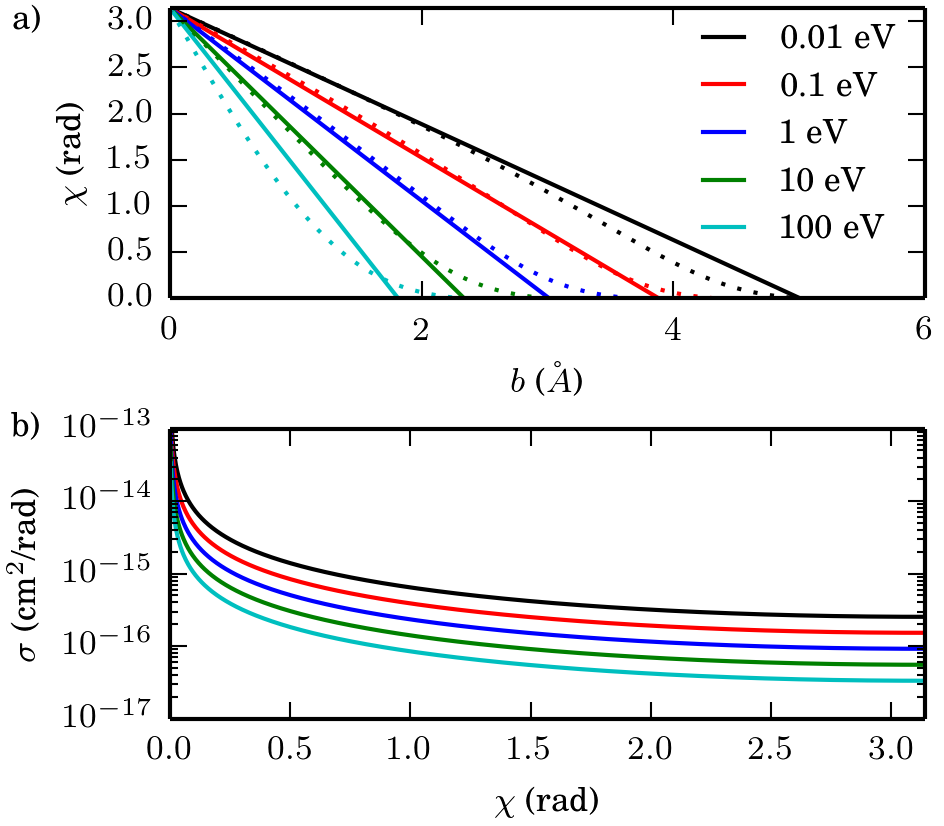}
	}
	\caption{a) Scattering angle $\chi$ of Ar-Al 
	collisions plotted as a function of the impact 
	parameter $b$ for the linear dependence of the M1 model 
	(solid lines) and the corresponding analytic calculations 
	(dotted lines) for different interaction energies. 
	b) The corresponding Ar-Al differential 
	collision cross-sections based on 
	eq.~(\ref{eq:differential_cross_section}).}
	\label{fig:combined_impact_parameter_differential_cross_section}
\end{figure}

Of major importance for this work are combined collisions between argon and aluminum. For the analytic interaction potential as well as for the M1 model the corresponding collision parameters can be evaluated based on the parameters specified for the individual species. For the analytic interaction potential a combining rule $\Phi(r) = \sqrt{\Phi_\textrm{Ar}(r)\Phi_\textrm{Al}(r)}$ is used \cite{hirschfelder_molecular_1964, abrahamson_born-mayer-type_1969}. For the M1 model a linear interpolation $d_\textrm{ref,Ar-Al} = (d_\textrm{ref,Ar} + d_\textrm{ref,Al})/2$ and $\omega_\textrm{Ar-Al} = (\omega_\textrm{Ar} + \omega_\textrm{Al})/2$ is performed \cite{bird_molecular_1994}. The corresponding results obtained from the analytic interaction potential as well as the M1 dependence for Ar-Al collisions are displayed in Fig.~\ref{fig:combined_impact_parameter_differential_cross_section}a. It is important to notice that also in the combined case the scattering characteristics are captured well -- within the limitations of the individual argon/aluminum fits.

The differential cross-section for an Ar-Al collision as calculated from the scattering dependence of the M1 model \cite{robinson_energetic_1979}
\begin{align}
	\sigma(\chi) = \dfrac{b}{\sin\chi} 
	\left| \dfrac{db}{d\chi} \right| = \dfrac{d_\textrm{i}^{2}(V_\textrm{r})}{\pi\sin\chi} 
	\left( 1 - \dfrac{\chi}{\pi} \right)
	\label{eq:differential_cross_section}
\end{align}
is shown in Fig.~\ref{fig:combined_impact_parameter_differential_cross_section}b. Its angular dependence forthrightly reflects the anisotropy of the underlying collision model (specifically, there is no symmetry about $\pi/2$). For glancing collisions (i.e., $b \rightarrow d_\textrm{i}(V_\textrm{r})$ and thus $\chi \rightarrow 0$) the differential collision cross-section diverges. This stems from $\sin\chi$ in the denominator and reflects the vanishing of the differential $d\Omega = \sin\chi d\chi d\psi$. The singularity at $\chi = \pi$ is canceled. Notably, integrals over $\sigma(\chi)d\Omega$ remain finite. In consequence the total collision cross-section $\sigma_\textrm{T}$, the momentum (or viscosity) cross-section $\sigma_\textrm{M}$, and the diffusion cross-section $\sigma_\textrm{D}$ can be obtained \cite{vincenti_introduction_1967, kersch_transport_2011}. The total collision cross-section which enters into eqs.~(\ref{eq:collision_probability_basic}) to (\ref{eq:collision_candidates}) finally reads $\sigma_\textrm{T} = \pi d_\textrm{i}^{2}(V_\textrm{r})$.

\subsection{Sputtered Particle Distribution Function}\label{sec:thompson}

\begin{figure}[t]
	\centering
	\resizebox{8cm}{!}{
		\begin{tabular}{c}
			\includegraphics{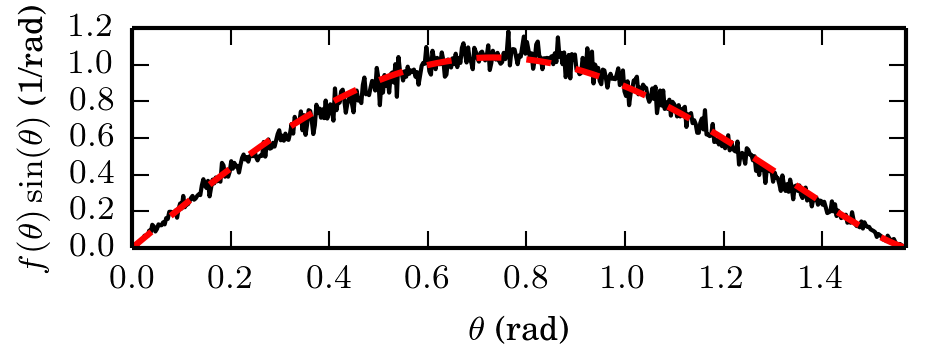}
		\end{tabular}
	}
	\caption{Angular distribution of sputtered aluminum obtained from 
	TRIDYN simulations (solid black line) by Bienholz et al.\ 
	\cite{bienholz_multiple_2013} and analytic fit using a 
	cosine distribution eq.~(\ref{eq:cosine}) (dashed red line).}
	\label{fig:sputtered_ADF}
\end{figure}

\noindent The angular and energy distribution of particles sputtered of various kinds of surfaces has undergone extensive analyses in the past decades starting from the work of Thompson and Sigmund \cite{thompson_ii._1968, sigmund_theory_1969} and including numerous other works \cite{wehner_angular_1960, vossen_sputtering_1974, moller_tridyn_1984, betz_energy_1994, goehlich_angular_2000, stepanova_estimates_2001}. In this paper a rather empirical fitting approach is pursued where the angular and energy dependences of the respective distributions are compared and fitted to results reported by Bienholz et al.\ obtained from TRIDYN simulations \cite{bienholz_multiple_2013}. Instead of directly using their Monte Carlo results, analytic expressions are used in the TMPM simulations to initially specify the sputtered particle velocities. For this purpose, the normalized flux distribution of sputtered particles may be denoted by \cite{nanbu_synthetic_2013}
\begin{align}
	f(E,\theta,\varphi) dE d\Omega = f(E|\theta) dE 
	\times f(\theta) \sin\theta d\theta \times f(\varphi) d\varphi,
\end{align}
with the solid angle $d\Omega = \sin\theta d\theta d\varphi$. $\varphi$ is the uniformly distributed azimuthal emission angle with $f(\varphi) = (2\pi)^{-1}$. The energy distribution $f(E|\theta)$ is assumed to be dependent on the polar angle of emission $\theta$ which itself is measured from the surface normal ($0 \leq \theta \leq \pi/2$). For the latter a cosine-law \cite{stepanova_estimates_2001, greenwood_correct_2002} is assumed with
\begin{align}
	f(\theta) = B \cos^{\alpha^2}\left( 
	\theta \right).\label{eq:cosine}
\end{align}
$B$ is a normalization factor such that $\int_0^{\pi/2} f(\theta) \sin\theta d\theta = 1$. $\alpha^2 > 0$ is a parameter shaping the distribution to be over or under-cosine, respectively. In order to capture the angular distribution obtained by Bienholz et al.\ $\alpha = 1.1111$ was fitted making the distribution slightly over-cosine \cite{wehner_angular_1960, vossen_sputtering_1974}. The corresponding distribution is displayed in figure~\ref{fig:sputtered_ADF}. The excellent agreement of eq.~(\ref{eq:cosine}) with the Monte Carlo data is evident.

As concerns the particles' energy distribution often the angle-independent Thompson distribution is used \cite{thompson_ii._1968}. It is, however, only a rough estimate as observed in experiments and simulations \cite{betz_energy_1994, goehlich_angular_2000}. Based on the work of Stepanova and Dew \cite{stepanova_estimates_2001} an angle-resolved energy distribution
\begin{align}
f(E|\theta) = C(\theta) \dfrac{E}{(E+U)^{3-2m}} 
\left[ 1 - \left( \dfrac{U+E}{U+\Lambda E_\textrm{inc}} 
\right)^n \right]\nonumber\\ 
\times \exp\left[ -A \left( \dfrac{m_\textrm{i}}
{m_\textrm{t}}\dfrac{U + E s_{\theta}^q(\theta)}{E_\textrm{inc}} 
\right) ^Q\right]\label{eq:thompson}
\end{align}
can be utilized -- that is, the conditional probability given $\theta$. The angular dependence is specified by the shape function $s_{\theta}(\theta) = \cos\theta$, which is clearly asymmetric with respect to $\pi/4$. The given angular dependence strictly only holds for ions under normal incidence. The case of ions under oblique incidence is a different matter \cite{wehner_angular_1960, goehlich_angular_2000}. This topic is beyond the scope of this work, as the assumption of normal incidence is well justified due to a high sheath potential and a low collisionality for the situation of interest. The parameters specifying the distribution of eq.~(\ref{eq:thompson}) can be divided into two groups: i) the rather empirical parameters $m=0.212$, $\Lambda=0.14$, $n=0.5$, $A=13$, $q=2-m_\textrm{t}/(4 m_\textrm{i})$, and $Q=0.55$ which govern the functional shape of the distribution. These parameters are specified in \cite{stepanova_estimates_2001} and/or fitted to the TRIDYN results. ii) A set of parameters which are physically justified. That is, $U = E_\textrm{b,Al}=3.36$~eV is commonly chosen as the binding energy, the energy of ions incident on the target is $E_\textrm{inc} = e V_\textrm{SB}$, the incident ion mass is $m_\textrm{i}=m_\textrm{Ar}$ and the target atomic mass is $m_\textrm{t}=m_\textrm{Al}$. $C(\theta)$ is again used for normalization and chosen such that $\int_0^\infty f(E|\theta) dE = 1$.

\begin{figure}[t]
	\centering
	\resizebox{8cm}{!}{
		\includegraphics{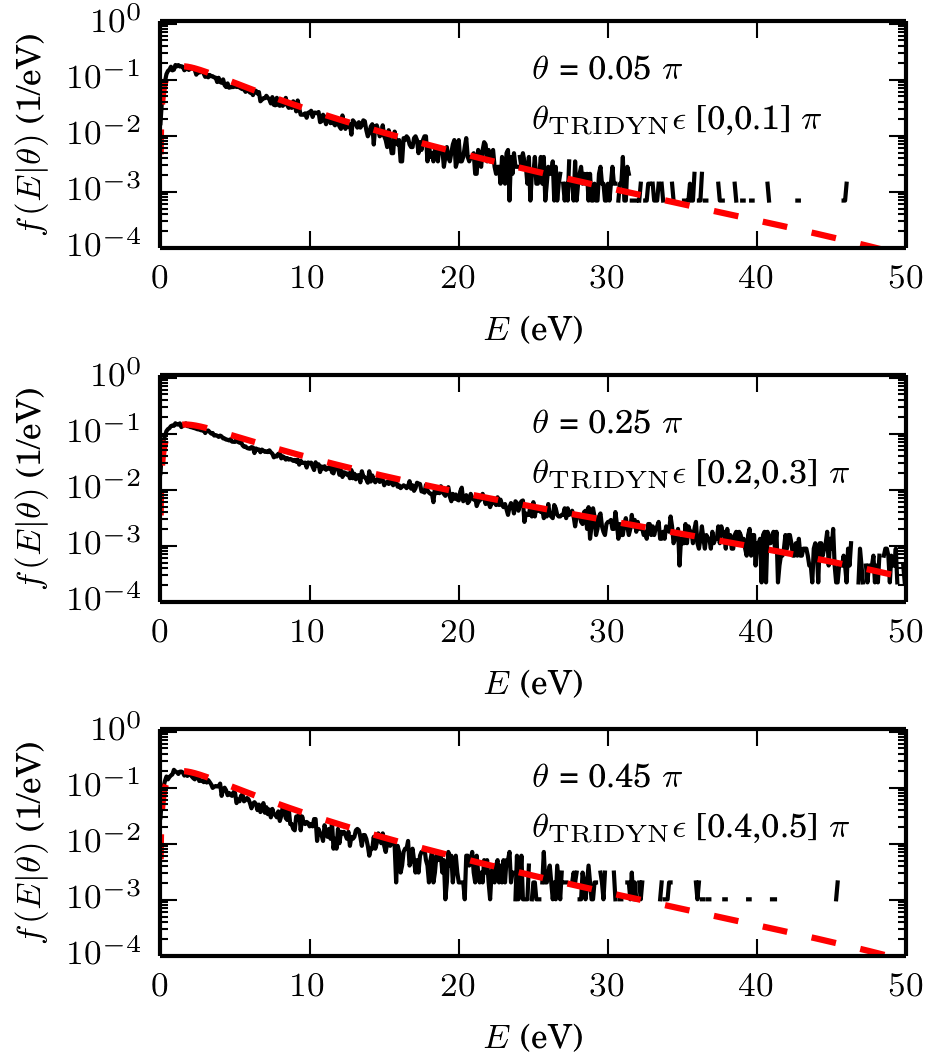}
	}
	\caption{Energy distribution of sputtered aluminum obtained via 
	TRIDYN simulations (solid black lines) by Bienholz et al.\ 
	\cite{bienholz_multiple_2013} and analytic fit using a 
	modified Thompson distribution eq.~(\ref{eq:thompson}) 
	(dashed red lines) for different emission angles $\theta$.}
	\label{fig:sputtered_EDF}
\end{figure}

The expression referenced for the angle-dependent energy distribution $f(E|\theta)$ seems to be appropriate for most sputtering applications \cite{goehlich_angular_2000, stepanova_estimates_2001, nanbu_synthetic_2013}. However, when comparing the TRIDYN results to the proposed expression, the angular dependence is captured well only for emission angles $\theta \leq \pi/4$. Above $\theta \approx \pi/4$, for the Monte Carlo results the high energy tail of the distribution drops, while the $\cos\theta$-dependence in the original expression yields a monotonic rise of the latter. To better reflect the Monte Carlo data, instead, a shape function $s_{\theta}(\theta) = 1 - k \sin(2\theta)$ is suggested. Its symmetry about $\pi/4$ allows to better capture the observed angular dependence, where $k = 0.5$ determines its strength. It is important to note that due to the choice of the exponents $q$ and $Q$ the symmetry about $\pi/4$ is broken, despite the symmetric choice of $s_{\theta}(\theta)$. The angle-resolved energy distribution $f(E|\theta)$ together with the Monte Carlo results of Bienholz et al.\ is depicted in figure~\ref{fig:sputtered_EDF} for emission angles $\theta=\left\{ 0, \pi/4, \pi/2 \right\}$. As most evident from the high energy tail, for emission angles larger than $\theta \geq \pi/4$ the distribution drops again. This is well captured by the modified shape function. Even for the improved angular dependence some deviations are observed. In particular for energies from a few up to $E \leq 30$~eV the analytic expression slightly overestimates the Monte Carlo results. The overall trend however is captured much better, compared to a monotonically rising high energy tail assumed for the original shape function or the classical angle-independent Thompson distribution. Without going into detail it can be stated that the specific choice of the shape function indeed has a significant influence for the cases of interest in this work.

From a numerical perspective the velocity distribution according to eqs.~(\ref{eq:cosine}) and (\ref{eq:thompson}) has to be assigned to individual particles entering the simulation domain. The angular distribution is specified by the polar angle $\theta = \arccos( R_f^{\beta} )$ with $\beta = 1/(\alpha^2+1)$ \cite{greenwood_correct_2002}, and the azimuthal angle $\varphi = 2 \pi R_f$. $R_f \in \left[0,1\right]$ are again independent, uniformly distributed random numbers. To sample the energy distribution the acceptance-rejection method is employed in accordance with reference \cite{nanbu_synthetic_2013}.

\section{Results and Discussion}\label{sec:results}

\subsection{Three Dimensional Configuration}\label{sec:three_dimensional}

\begin{figure}[t]
	\centering
	\resizebox{8cm}{!}{\includegraphics{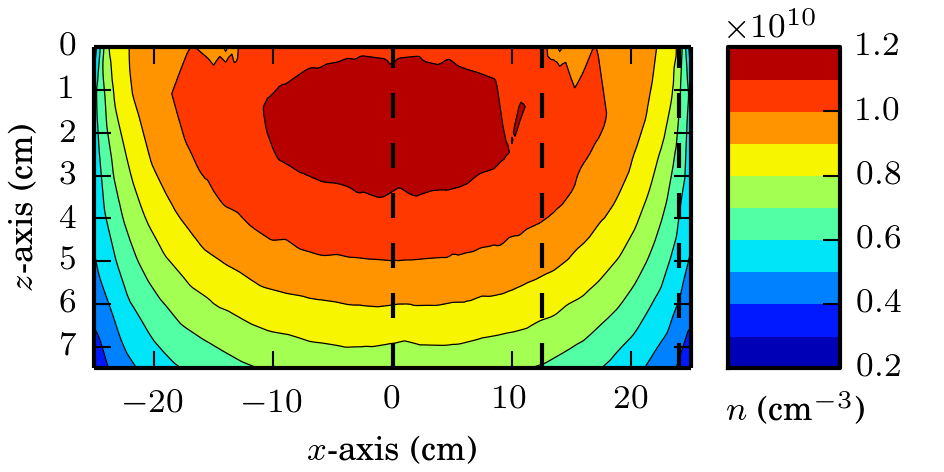}}
	\caption{Sectional drawing of the reactor chamber zoomed 
	to the electrode gap region indicated by the dashed red frame 
	in figure~\ref{fig:geometry}. The aluminum density $n$ is 
	displayed. The dashed black lines indicate the positions to be considered later in figure~\ref{fig:3D_local_density_velocity}.}
	\label{fig:3D_slice_density}
\end{figure}

\noindent The geometry and the discharge parameters from section~\ref{sec:setup} specify the reference case of interest for the analysis of the sputtered particle transport within the given MFCCP configuration. For this purpose the model described in the foregoing section~\ref{sec:numerical_method} has been employed. First, the complete three dimensional geometry is considered. Figure~\ref{fig:3D_slice_density} presents a cross-sectional slice through the reactor geometry zoomed to the gap region between the two electrodes as indicated in figure~\ref{fig:geometry}. The simulated aluminum density $n$ is plotted. The target is at the top at $z=0$~cm, while the opposing electrode (substrate) is at $z=7.5$~cm. Due to complete adsorption of the aluminum neutrals at the surfaces, the latter essentially act as particle sinks. On the contrary, sputtering takes place at the target surface only (i.e., resputtering from the substrate is neglected). In longitudinal  direction $z$ the aluminum density exhibits an asymmetric profile with a distinct maximum on the axis of symmetry, $z \approx 1.85$~cm from the target. This is on the order of a mean free path which under the assumption of a Maxwellian velocity distribution at $T=650$~K (likely underestimating the actual mean free path for the non-equilibrium situation) is approximately $\lambda_\textrm{mfp} \approx \left( n_\textrm{bg} \pi d_\textrm{ref,Ar}^2 \sqrt{2} \right)^{-1} \approx 1.9$~cm. This distinct maximum is an expected feature, as previously reported \cite{turner_monte_1989, clenet_experimental_1999}. While all surfaces act as particle sinks, only the target surface also acts as a particle source. In consequence, the density is minimal directly at the surfaces (in particular the electrodes). The density at the target exceeds the density at the substrate. A local maximum is observed in the electrode gap between. For higher pressures the maximum is more pronounced in the center and shifts towards the target as the pressure is increased.

The most significant observation from figure~\ref{fig:3D_slice_density} is that the density profile in the electrode gap is laterally inhomogeneous, despite the assumption of a homogeneous particle flux imposed over the target surface. The radial dependence is indeed mainly governed by the exact geometry as well as the imposed flux. More precisely, it is governed by the initial distribution of sputtered particles (that is, the spread of the polar emission angle which stems from the cosine distribution) as well as the lateral distribution of the imposed flux (viz., homogeneous here). More importantly, however, it is influenced by the chamber volume outside the electrode gap. Its role as a particle sink or source changes depending on the degree of backscattering and the distance to the outside (absorbing) walls: For low pressures few particles are backscattered and the volume acts predominantly as a particle sink; for higher pressures with increasing backscattering the volume to some degree is a particle source. A substantial flux of particles may reenter the gap region. The problem can be fundamentally viewed as follows: In principle particles have a chance to be lost to any wall (particle sink) inside or outside the electrode gap region. Yet, particles that have already traversed to the outside region also have a chance of being inwardly scattered. Even though it is considerably smaller than the chance of traversing to any outside wall it remains finite. This effect is naturally amplified by the nearly circular geometry. The flux of particles leaving (entering) the electrode gap region through the girthed area $A_\textrm{girthed} = \pi W D \approx 1178~\textrm{cm}^2$ is proportional to the inside (outside) particle density. As the net flux of particles is outwardly directed, a gradient in the density inherently establishes. Consequently, the density significantly drops in radial direction.

\begin{figure}[t]
	\centering
	\resizebox{8cm}{!}{\includegraphics{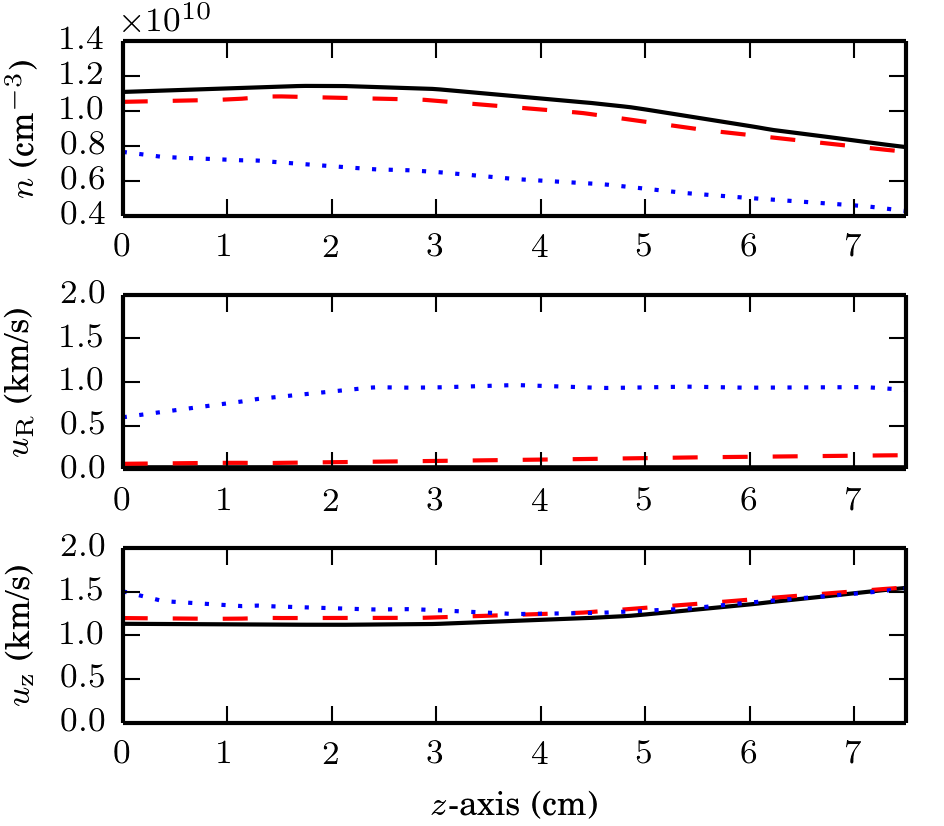}}
	\caption{Aluminum density $n$ (above), mean radial velocity 
	$u_\textrm{R}$ (center), and mean longitudinal velocity 
	$u_\textrm{z}$ (below) plotted over the electrode gap 
	$z$ for different radial positions $R=0, 12.5 \textrm{ and } 24$ 
	cm (solid black, dashed red and dotted blue line, respectively).}
	\label{fig:3D_local_density_velocity}
\end{figure}

The observed density gradient from the center towards the electrode edge is illustrated by the density and mean velocity profiles plotted over the electrode inter-spacing for different radial positions as given in figure~\ref{fig:3D_local_density_velocity}. For three different radial positions (as indicated in figure~\ref{fig:3D_slice_density} by vertical dashed lines) one can clearly observe a drop of the aluminum density towards the electrode edge. This is accompanied by a substantial increase in the overall mean flow velocity. The density is relatively homogeneous for $R \leq 13$~cm, which is in good agreement with the experimental observation of a homogeneous deposition rate on the substrate for radii $R \leq 12$~cm \cite{bienholz_kapazitiv_2014}. The drop in density towards the edge of the electrode ranges from a factor of 1.5 to 2 (between the center and the edges) depending on the distance from the target. As argued earlier, this gradient manifests in a substantial radial particle flow. While the longitudinal (i.e., $u_\textrm{z}$) component remains almost constant, there is a substantial increase in $u_\textrm{R}$. By symmetry, the net flow of sputtered particles is directed out of the electrode gap region. Note that the flux of particles is conserved (that is, no particle is lost on its way).

\begin{figure}[t]
	\centering
	\resizebox{8cm}{!}{\includegraphics{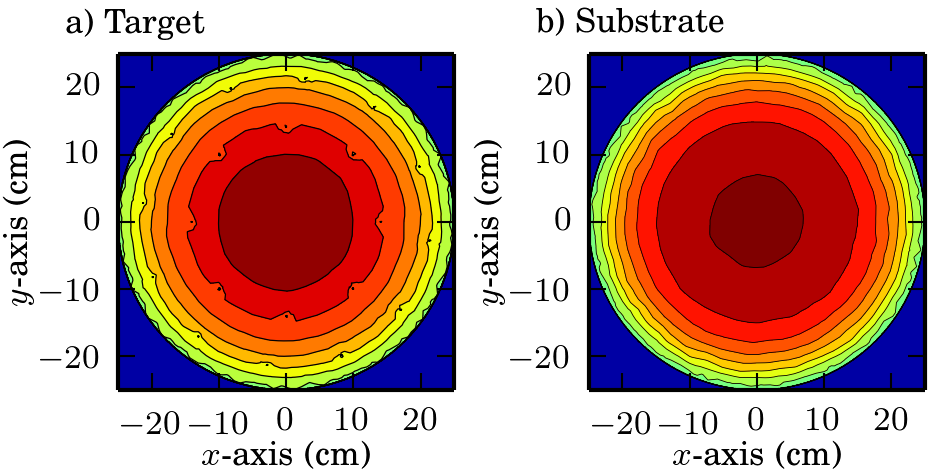}}
	\caption{Lateral distribution of the absorbed fluxes of aluminum 
	onto a) the target and b) the substrate electrode. The color 
	range is from zero to a) $2.953 \times 10^{14} 
	\textrm{ cm}^{-2}\textrm{s}^{-1}$ and 
	b) $1.214 \times 10^{15} \textrm{ cm}^{-2}\textrm{s}^{-1}$.}
	\label{fig:3D_lateral_density}
\end{figure}

The lateral distribution of the particle flux onto the substrate, respectively onto the target is another measure for the characterization of the afore mentioned geometric influence. Note that what is shown in figure~\ref{fig:3D_lateral_density} is the flux of particles lost to the surfaces only. It is not super-imposed with the source flux originating from the target. Due to the azimuthal symmetry of the reactor chamber being disturbed only outside the narrow electrode gap region, the flux distribution at the substrate is also nearly azimuthally symmetric. (The symmetry is in addition slightly disturbed by a number of small gas inlets embedded into the target.) The radial profile essentially projects what was observed for the density. Moreover, now the amount of sputtered as well as deposited material can be quantified as follows: The imposed source flux at the target integrated over the area is $\dot{N} = \int_{\textrm{A}} dA~\Gamma_\textrm{0} = 1.6 \times 10^{15} \textrm{ cm}^{-2}\textrm{s}^{-1} \times 1963 \textrm{ cm}^{2} = 3.14 \times 10^{18} \textrm{ s}^{-1}$. On the other hand, the number of particles absorbed by the substrate electrode is $\dot{N}_\textrm{s} = 1.95 \times 10^{18} \textrm{ s}^{-1}$, while for the target itself it is $\dot{N}_\textrm{t} = 4.61 \times 10^{17} \textrm{ s}^{-1}$. Consequently, a fraction of $62.1 ~\%$ of the original particle flux reaches the substrate, while a fraction of $14.7 ~\%$ is reflected back and reattaches to the target, and $23.2 ~\%$ is lost to the chamber walls. The latter can be easily verified by comparison with the effective particle loss areas. The total surface area of the electrode gap region is $A_\textrm{gap} = 2 A_\textrm{target} + A_\textrm{girthed} \approx 5104~\textrm{cm}^2$. The relative contribution of the girthed area then is $A_\textrm{girthed} / A_\textrm{gap} \approx 23~\%$. This number is in remarkable agreement with the flux fraction lost to the chamber walls, although this consideration does not account for the directionality of the sputtered particles as well as the additional influx of backscattered particles. The former manifests in the imbalance between the number of particles reaching the target and substrate surfaces, respectively (i.e., it is not $A_\textrm{target} / A_\textrm{gap} \approx 38~\%$ each).

The observed deposition probabilities are moreover consistent with experimental observations of Rossnagel \cite{rossnagel_deposition_1988} who investigated the deposition probability in a cylindrically symmetric magnetron discharge under comparable conditions. For an aluminum target it was found that with a target-to-substrate distance of $D=5$~cm and an argon pressure of $p=0.7$~Pa (thus $pD=3.5$~cm~Pa is comparable to $pD=3.75$~cm~Pa in this work) the deposition probability was $60~\%$ for the opposing sample plane, $12~\%$ for the magnetron plane and $10~\%$ for the side areas. The first two values are in excellent agreement considering the given differences. On the other hand, the discrepancy in the side wall contribution is simply justified by the fact that in the work of Rossnagel the side walls do not comprise of all of the remaining chamber walls (as compared to this work). The proposed results are moreover in reasonable agreement with the work of Turner et al.\ \cite{turner_monte_1992} who determined the deposited fraction of sputtered silicon and sodium ranging between $60 ~\%$ and $75 ~\%$ (the fraction for aluminum can only be roughly estimated based on the respective binding energy and atomic mass).

The longitudinal profiles of the density and mean velocity along the gap (in $z$ direction) are a different matter. Over all radii the longitudinal mean velocity component ($u_\textrm{z}$) increases slightly by a factor of about 1.4 from the target to the substrate electrode, while the radial mean velocity component is nearly constant (unless very close to the edge of the target circumference). This indicates two aspects: Firstly, as previously reasoned, the geometry has a large influence on the detailed behavior of the particle fluxes and therefore the deposition itself (in particular laterally). Secondly, the drop in density along the $z$-axis is mainly governed by a one dimensional transport process where particles interact with the background gas (i.e., scattering and thus slowing down of particles). The first aspect is important for reliable predictions of the sputtering and deposition as concerns specific details of the geometry. On the other hand, the second aspect is equally important as it indicates the principle mechanism of the particle transport from the target to the substrate.

\begin{figure}[t]
	\centering
	\resizebox{8cm}{!}{\includegraphics{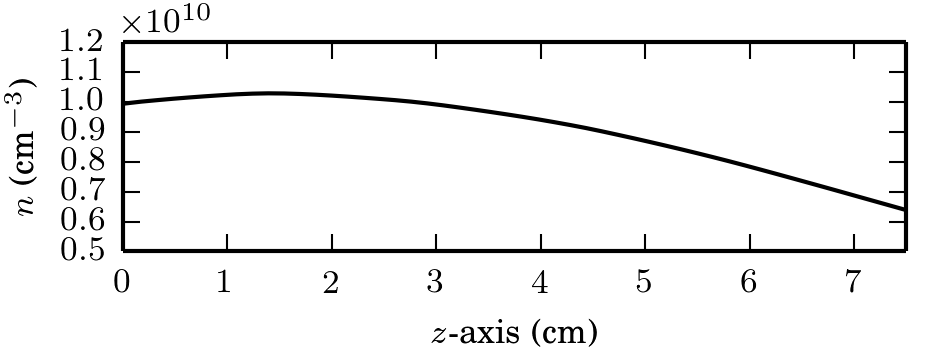}}
	\caption{Line-of-sight averaged aluminum density $n$ obtained from figure~\ref{fig:3D_slice_density} plotted as a function of the distance $z$.}
	\label{fig:3D_line_of_sight_density}
\end{figure}

The hypothesis of a mainly one dimensional transport mechanism can also be argued on the basis of the laterally averaged aluminum density profile. By integrating the cross-sectional density data displayed in figure~\ref{fig:3D_slice_density} along the $x$-axis (cf., Fig.~\ref{fig:3D_slice_density}), the line-averaged aluminum density profile depicted in figure~\ref{fig:3D_line_of_sight_density} (plotted as a function of the gap distance $z$) is obtained. Of course due to the averaging all radial information is lost. It is seen that the profile shape of the averaged density is governed by the homogeneous central region of the discharge. As evident from the spatially resolved density (cf., Figs.~\ref{fig:3D_slice_density} and \ref{fig:3D_local_density_velocity}) the latter varies from a round shape, which is maximum at some finite value $1.5 < z < 1.85$~cm for $R < 13$~cm (with a min/max ratio of roughly $0.79 \times 10^{10}~\textrm{cm}^{-3} / 1.15 \times 10^{10}~\textrm{cm}^{-3} = 69~\%$), to a nearly linear trend at $R = 24$~cm. Although the averaged profile is slightly smoothed the principle curvature remains round. The governing transport mechanism is unaltered by the averaging. The averaged profile has its maximum somewhere in the electrode gap (depending on pressure; here at $z \approx 1.39$~cm, roughly $25~\%$ less than non-averaged). Of interest in this context is the relative change in values. The averaged density ranges from a maximum value of $n \approx 1.03 \times 10^{10} \textrm{ cm}^{-3}$ at $z \approx 1.39$~cm down to $n \approx 0.64 \times 10^{10} \textrm{ cm}^{-3}$ at the substrate electrode. This is a relative drop to about $62~\%$ the maximum value (slightly less than non-averaged). On the other hand, towards the target it only drops down to $n \approx 0.98 \times 10^{10} \textrm{ cm}^{-3}$, approximately $95~\%$ the maximum value. Note that the obtained averaged density profile cannot straight forwardly be compared to experimental data from optical emission spectroscopy (OES) measurements as these stem from a line-of-sight integration over the optical acceptance cone of the apparatus \cite{bienholz_multiple_2013}. In order to compare our spatially resolved model results with the experimental observations, the calculated data would have to undergo a similar analysis. A more detailed comparison between experiments and simulations focusing on the radial dependence and the three dimensional features of the transport of sputtered particles will be the topic of future work.

\subsection{One Dimensional Analysis}\label{sec:one_dimensional}

\noindent As previously reasoned, the governing transport mechanisms inside the MFCCP are of mainly one dimensional nature (c.f., section~\ref{sec:three_dimensional}). To underpin the previously said, the analysis can be significantly facilitated by making use of a simplified one dimensional transport model. The previously applied model can be straightforwardly employed. Notably, the dimensional reduction leads to a welcomed improvement in performance (runtime of hours instead of days). The major achievement, however, lies within the much simplified analysis. Individual particles are now described in a continuous 1D-3V phase space (i.e., three velocity components are maintained but their position is restricted to the $z$ direction along the gap). The density and mean velocity, therefore, depend on $z$ only. The vectorial net flux is solely longitudinal ($\vec{\Gamma} = n u_\textrm{z} \vec{e}_\textrm{z}$). As verified later, this is a consequence of the azimuthal symmetry of the velocity distribution of the Lagrangian particles.

\begin{figure}[t!]
	\centering
	\resizebox{8cm}{!}{\includegraphics{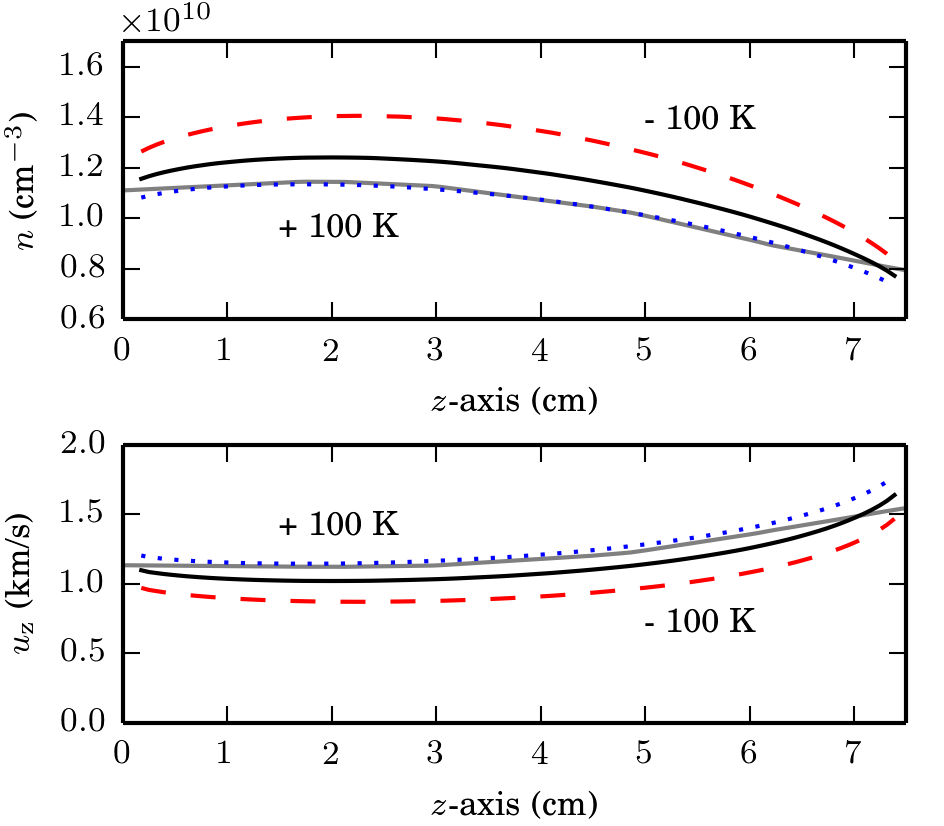}}
	\caption{Aluminum density $n$ (top) and mean longitudinal 	velocity $u_\textrm{z}$ (below) obtained for the one 	dimensional case plotted as a function of $z$. The background temperature is varied: $T=550$~K (dashed red), $T=650$~K (solid black) and $T=750$~K (dotted blue). The 3D result from Fig.~\ref{fig:3D_local_density_velocity} ($R=0$~cm) is given for reference (solid gray).}
	\label{fig:1D_temperature_variation_density}
\end{figure}

Analogous to the previous analysis, the aluminum density as well as the longitudinal mean velocity (now equal to the magnitude of the overall mean velocity) are given in figure~\ref{fig:1D_temperature_variation_density} for the one dimensional situation. More precisely what is shown are results obtained for the reference case of section~\ref{sec:setup} with $T=650$~K (solid black line) and for a variation of the background temperature $T = 650 \textrm{ K } \pm 100$~K (twice the confidence interval of the temperature measurement of Bienholz et al.\ \cite{bienholz_multiple_2013}). A pressure of $p = 0.5$~Pa is maintained, so the corresponding background densities are $n_\textrm{bg} = \{ 6.585,~5.572,~4.829 \} \times 10^{13} \textrm{ cm}^{-3}$. The reduction (increase) of the temperature has two effects: On the one hand, the mean free path is decreased (increased), on the other hand thermalization strives towards a lower (higher) equilibrium value. For a reduced temperature of $T = 550$~K (dashed red lines) and a correspondingly larger background density the slowing down of energetic sputtered particles is more efficient. Consequently, the mean velocity reduces by roughly $- 10~\%$. The aluminum density increases (by the same amount) due to flux conservation and thus accumulation of slow particles. More importantly, the overall behavior remains the same. That is, the density profile still exhibits a pronounced maximum of $n \approx 1.4 \times 10^{10} \textrm{ cm}^{-3}$ now at $z \approx 2.3$~cm. The contrary behavior is seen when the collisionality is decreased by raising the temperature to $T = 750$~K (dotted blue lines). The aluminum density decreases by roughly $- 10~\%$ because less and faster particles accumulate in the volume due to fewer collisions. (The mean velocity increases accordingly.) Also in this case the generic profile of both the density and the mean velocity remain similar with a maximum in density at $z \approx 1.8$~cm.

As previously reported \cite{turner_monte_1989, clenet_experimental_1999}, the collisionality has a strong influence on the strength of the local maximum in density (for different pressures at a constant temperature). In the present case the relative strength of the maximum in aluminum density (absolute max/min) ranges from approximately 1.75 at $T = 550$~K down to 1.38 for $T = 750$~K. With the same reasoning this is merely a consequence of the variation in collisionality rather than the temperature.

It is instructive to take a closer look at the reference case of $T = 650$~K for a detailed understanding of the underlying transport (Fig.~\ref{fig:1D_temperature_variation_density}, solid black lines). The principle shape of the 1D aluminum density profile is in fact much the same as for the 3D case. A pronounced maximum of $n \approx 1.24 \times 10^{10} \textrm{ cm}^{-3}$ is observed approximately $z \approx 2$~cm from the target (comparable to $n \approx 1.15 \times 10^{10} \textrm{ cm}^{-3}$ at $z \approx 1.85$~cm for the 3D results on the axis of symmetry). This is again a consequence of a balancing of the particle fluxes originating from the electrode surfaces (cf., sink/source). The nearly 8~\% increase is from the additional contribution of particles which are not ``lost'' to the girthed area. This balancing is also reflected by the observance of a higher density $n \approx 1.15 \times 10^{10} \textrm{ cm}^{-3}$ at the target, compared to $n \approx 0.75 \times 10^{10} \textrm{ cm}^{-3}$ at the substrate. It is worthwhile to note that the relative drop from the maximum value down to $60~\%$ at the substrate electrode is quite close to the drop observed for the 3D situation ($69~\%$ on the axis of symmetry; $62~\%$ averaged). This supports the hypothesis of a similarly governed transport mechanism in both situations.

The line of arguments is slightly more complicated regarding the profile of the mean velocity. Two hypothetical however short explanations can be argued:

\begin{itemize}
\item[i)] A decrease of the mean velocity from the target towards the substrate may be expected simply because the slowing down of particles scales with the number of collisions on the particles' paths (governed by the mean free path). The density and mean velocity are balanced by the steady-state continuity equation $\nabla \cdot \left(n \vec{u} \right) = 0$ (i.e., the flux of particles has to be conserved). An increase in density is predicted, which is not observed.

\item[ii)] Based on the continuity equation and given the density distribution along the gap it can be argued that the mean velocity is fully specified by the ratio of the imposed flux and the aluminum density with $\vec{u} = \vec{\Gamma}_\textrm{0} / n$. This in principle gives the correct profile, however, scaled by the ratio of the imposed to the net particle flux. 
\end{itemize}

The actual behavior is in fact the same for the 1D and 3D situation: The net flux of particles is a superposition of the predefined (or imposed) flux emitted from the target $\Gamma_\textrm{0}$ and the flux of backscattered particles traversing in opposite direction and reattaching to the target. In the 1D case, this flux of backscattered particles can be quantified to $\Gamma_\textrm{t} = 3.337 \times 10^{14} \textrm{ cm}^{-2}\textrm{s}^{-1}$, which makes about $21 ~\%$ of the initially imposed flux. Correspondingly the flux of particles that reach the substrate is $\Gamma_\textrm{s} = 1.266 \times 10^{15} \textrm{ cm}^{-2}\textrm{s}^{-1}$, equal to a deposition probability of $79 ~\%$. Expectedly both numbers are slightly higher compared to the 3D case, because now particles can either be lost to the target or the substrate ($\Gamma_\textrm{0} = \Gamma_\textrm{t}+\Gamma_\textrm{s}$). The ratio of the fluxes to the electrodes, however, is comparable: $\Gamma_\textrm{t}/\Gamma_\textrm{s} |_\textrm{1D} = 0.26$ vs. $\Gamma_\textrm{t}/\Gamma_\textrm{s} |_\textrm{3D} = 0.24$. Now, to explain the mean velocity rise two aspects are important: Firstly, the substrate strictly acts as a particle or flux sink. The reversely directed flux originates from the volume due to backscattering (i.e., it vanishes directly at the substrate). Secondly, the density as well as the mean velocity are fully specified only by the sum of both fluxes. As both fluxes are of opposite sign they compensate each other. When super-imposed they are truly subtractive, not additive. (Note that the density as a scalar quantity indeed behaves truly additive.) This explains why the net flux as well as the mean velocity are overestimated from the previous argument.

\begin{figure}[t!]
	\centering
	\resizebox{8cm}{!}{\includegraphics{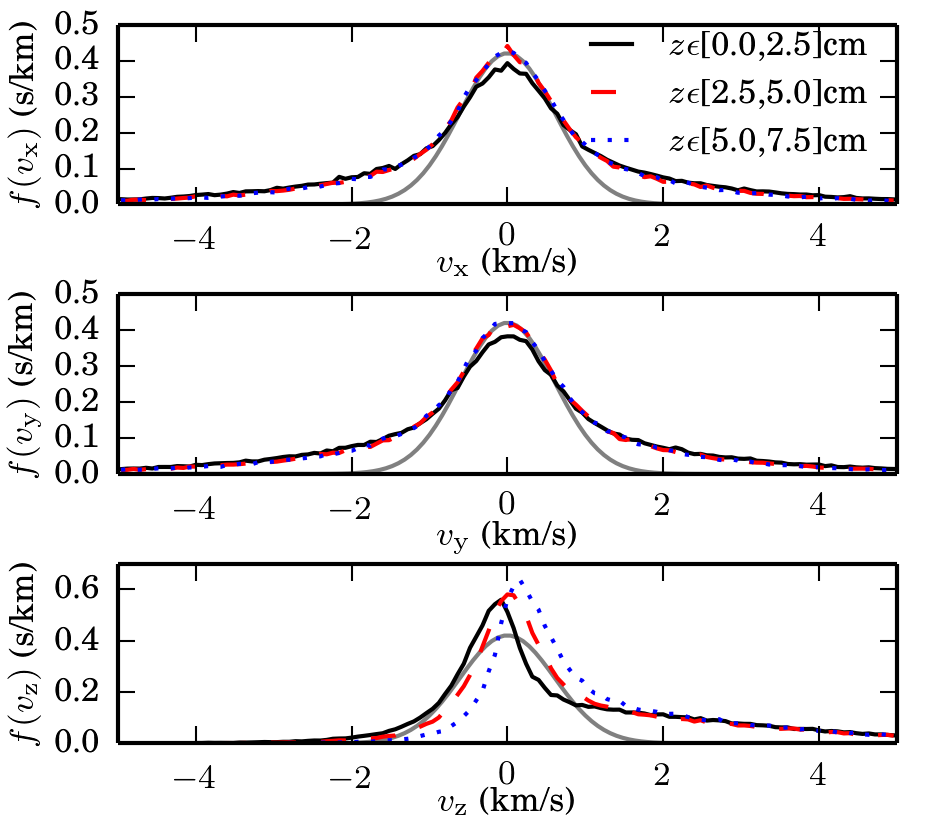}}
	\caption{Aluminum velocity distributions for the different 
	velocity components and a background temperature of 
	$T=650$~K at a pressure $p=0.5$~Pa. For comparison a 
	Maxwellian velocity distribution of temperature 
	$T = 1200$~K is plotted (solid gray line).}
	\label{fig:1D_sectioned_VDF}
\end{figure}

To prove this hypothesis it is instructive to investigate the underlying velocity distribution function (VDF) of the Lagrangian particles. Figure~\ref{fig:1D_sectioned_VDF} presents the three corresponding velocity distributions within three different regions inside the electrode gap. For comparison all distributions (in their respective spatial interval) are normalized such that $\int_{-\infty}^{\infty} f(v_\textrm{k}) dv_\textrm{k} = 1$. Note that the individual particle velocities are denoted by $v_\textrm{k}$ to avoid confusion with the mean velocity $u_\textrm{k}$. Certain aspects are of importance with respect to the particle transport:

\begin{itemize}
\item[i)]The simulation model is one dimensional so by symmetry the velocity distributions in lateral direction $f(v_\textrm{x})$ and $f(v_\textrm{y})$ are expected to be indistinguishable (within the statistical error). Consequently, all but the z-component of the mean velocity $u_\textrm{z}$ discussed earlier actually vanish. As evident from the two topmost subfigures of figure~\ref{fig:1D_sectioned_VDF} this is strictly satisfied. There is no net flux of particles in any other direction than $\pm z$.

\item[ii)] The cosine angular distribution imposed at the target boundary specifies the initial velocities with $v_\textrm{z} \geq 0$, but non-zero $v_\textrm{x}$ and $v_\textrm{y}$. The magnitudes follow from the high energy tail of the modified Thompson distribution. Compared to a Maxwellian distribution the lateral VDFs $f(v_\textrm{x})$ and $f(v_\textrm{y})$ exhibit slightly overpopulated tails (only a fraction of particles is thermalized). Indeed the VDFs of only the backscattered particles would be much better represented by a Maxwellian distribution, due to collisions with the background gas. This can be reasoned from the lateral VDFs for particles residing in different intervals within the electrode gap. Most of the particles that reside close to the target are truly original (they have undergone fewer collisions with the background), thus preserving the original shape of their VDFs. Consequently, the agreement with a Maxwellian distribution is worse: it is slightly broader (solid black line) compared to regions further away from the target (dashed red and dotted blue line). The similarity of the tail of the distributions in all three regions stems from the initial cosine angular distribution as well as the comparably small number of scattering events. Note that even the slow particles are far from equilibrium with the background gas with temperature $T=650$~K. In the steady situation their distribution can be approximated with a Maxwellian temperature of $T=1200$~K (solid gray line).

\item[iii)] The situation is different for $f(v_\textrm{z})$, the distribution of the longitudinal velocity component. The individual particle positions are constrained to one spatial dimension, while three particle velocity components are maintained. Consequently, the degrees of freedom for the mean quantities of the system are restricted to this direction (i.e., the $z$-direction). By symmetry only $n$ and $u_\textrm{z}$ are allowed to establish distinct profiles along the gap distance. The latter simultaneously allows for a substantial variation of $f(v_\textrm{z})$ for different regions inside the electrode gap. The observed distributions can be explained as follows: Close to the target (solid black line) the flux of newly injected ``original'' particles is largest. At the same time the flux of backscattered particles is largest too, because these originate from the whole gap region behind. Consequently, in this region the flux of most ``ideal'' Thompson distributed particles (positively directed) interferes with the flux of most-thermalized backscattered particles (negatively directed). Moreover, slow (scattered) particles reside in a given volume for a longer period of time than fast particles and therefore seemingly accumulate in the volume. The described phenomenon can be well observed in $f(v_\textrm{z})$. A substantial contribution of backscattered particles with $v_\textrm{z}<0$ is observed. At the same time, the part of the distribution with $v_\textrm{z}>0$ is the superposition of the ``original'' Thompson distribution (with its high energy tail) and the contribution of positively directed scattered particles. The latter contribute most to the slow particle component. Directly at the target surface, however, they actually vanish (as there is no volume to originate from). A contrary effect can be observed in proximity to the substrate (dotted blue line). There, the tail of the distribution is most ``distorted'' (although it is actually quite unmodified due to the low collisionality). The flux of positively directed unaltered particles is smallest, while its scattered equivalent is largest. The drop in the negatively directed contribution comes from the fact that no particle originates from the substrate (a particle sink). Its contribution to the flux is smallest (it vanishes directly at the surface, assuming no resputtering). This explains the increase of the mean velocity towards the substrate previously observed and not understood. It is an immediate result of the interference of the positively and negatively directed particle fluxes. The mean velocity increase essentially results from the particle scattering inside the gap volume.
\end{itemize}

\section{Conclusions}\label{sec:conclusion}

\noindent The goal of this work is to provide sound insight into the not well understood physics of particle transport in large area capacitive sputtering devices. Several novel aspects have been addressed: i) The no-time counter idea has been adopted to the test multi-particle method. ii) A consistent set of scattering parameters to be used with the M1 collision model has been obtained. iii) A modified angular dependence for the initial energy distribution of sputtered particles was proposed. iv) Finally, using the advocated numerical simulation the governing transport mechanism of sputtered particles within the MFCCP reactor has been explained. The analysis was initially performed for the fully resolved three dimensional geometry. Therein, a strong influence of the geometry on the lateral transport has been observed, having immediate influence on the spatial distribution of coating formation on the substrate, the walls and the target itself. With respect to the longitudinal transport it has been postulated that the peculiar essence of the particle transport can be readily pictured in one spatial dimension. To justify this hypothesis a one dimensional simulation model was utilized. It has been shown that despite the drastic simplification the dominant transport mechanism is retained, while allowing for a much easier interpretation of the fundamental aspects. On the basis of the spatially resolved velocity distributions of the individual particles the observed density and mean velocity profiles have been explained by the interference of the intrinsic fluxes of ``original'' and scattered particles. This result is particularly valuable as it provides an insightful interpretation of the underlying transport mechanism, which can be directly related to the three dimensional situation. An in-depth comparison of numerical simulations of the three dimensional situation with spatially resolved experimental measurements will be the subject of a future work.


\section*{Acknowledgments}

\noindent This work is supported by the German Research Foundation (DFG) in the frame of Collaborative Research Centre SFB/TRR 87. The authors thank S. Gallian, D. Kr\"uger, F. Schmidt, and R. P. Brinkmann from the Institute of Theoretical Electrical Engineering, Ruhr University Bochum for fruitful discussions and in particular S. Bienholz, S. Ries, N. Bibinov, H. Kellner, and P. Awakowicz from the Institute for Electrical Engineering and Plasma Technology, Ruhr University Bochum for providing the original CAD design and their TRIDYN data for reference.
	
\bibliography{references.bib}

\end{document}